\documentclass{article}

\usepackage{microtype}
\usepackage{graphicx}
\usepackage{subfigure}
\usepackage{booktabs} 
\usepackage{amsmath}

\usepackage{hyperref}

\usepackage{mlsys2025}

\makeatletter
\let\ifmarginsmessedwith\iffalse
\makeatother
\mlsystitlerunning{Eliminating Multi-GPU Performance Taxes: A Systems Approach to Efficient Distributed LLMs}

\usepackage{xcolor}   
\usepackage{ifthen}   

\newcommand{\todo}[1]{}
\newcommand{\mawad}[1]{}
\newcommand{\mosama}[1]{}
\newcommand{\karthik}[1]{}
\newcommand{\sumanth}[1]{}
\newcommand{\octa}[1]{}

\ifdefined \FINAL
\else
\renewcommand{\todo}[1]{\par\noindent{\color{red}\textbf{TODO:} #1}\par}
\renewcommand{\mawad}[1]{\par\noindent{\color{blue}\textbf{mawad:} #1}\par}
\renewcommand{\mosama}[1]{\noindent{\color{violet}\textbf{mosama:} #1}}
\renewcommand{\karthik}[1]{\par\noindent{\color{cyan}\textbf{Karthik:} #1}\par}
\renewcommand{\sumanth}[1]{\par\noindent{\color{green}\textbf{Sumanth:} #1}\par}
\renewcommand{\octa}[1]{\par\noindent{\color{olive}\textbf{Octa:} #1}\par}

\fi

\begin{document}

\twocolumn[
\mlsystitle{Eliminating Multi-GPU Performance Taxes: A Systems Approach to Efficient Distributed LLMs}



\mlsyssetsymbol{equal}{*}

\begin{mlsysauthorlist}
\mlsysauthor{Octavian Alexandru Trifan}{uci,equal}
\mlsysauthor{Karthik Sangaiah}{amd}
\mlsysauthor{Muhammad Awad}{amd}
\mlsysauthor{Muhammad Osama}{amd}
\mlsysauthor{Sumanth Gudaparthi}{amd}
\mlsysauthor{Alexandru Nicolau}{uci}
\mlsysauthor{Alexander Veidenbaum}{uci}
\mlsysauthor{Ganesh Dasika}{amd}

\end{mlsysauthorlist}

\mlsysaffiliation{uci}{University of California, Irvine}
\mlsysaffiliation{amd}{AMD Research and Advanced Development}

\mlsyscorrespondingauthor{Octavian Alexandru Trifan}{otrifan@uci.edu}

\mlsyskeywords{Machine Learning, MLSys}

\vskip 0.3in

\begin{abstract}
As large language models~(LLMs) continue to scale, their workloads increasingly rely on distributed execution across multiple GPUs. However, the conventional bulk synchronous parallel~(BSP) model used in such settings introduces significant performance inefficiencies. To characterize these bottlenecks, we introduce the ``Three Taxes'' (Bulk Synchronous, Inter-Kernel Data Locality, and Kernel Launch Overhead) as an analytical framework. We propose moving beyond the rigid BSP model to address key inefficiencies in distributed GPU execution. By exploiting libraries like Iris for Triton, we gain access to in-kernel communication primitives that enable the design of novel fine-grained programming patterns, offering greater flexibility and performance than traditional BSP-based approaches. These patterns systematically eliminate the three taxes by creating direct, tile-level producer-consumer pipelines and replacing global barriers with fine-grained dataflow synchronization. Applying this methodology to critical kernels, from the foundational All-Gather + general matrix multiplication operation to the complex Flash Decode algorithm, we observe a 10-20\% speedup in end-to-end latency over BSP-based approaches, establishing a more programmable and efficient paradigm for distributed LLM workloads.
\end{abstract}
]



\printAffiliationsAndNotice{\textsuperscript{*}Work performed at AMD} 

\section{Introduction}
\label{submission}

The scale of modern Artificial Intelligence models, particularly large language models~(LLMs), continues to grow at a rapid pace, both in parameter count and computational demand. State-of-the-art models now regularly surpass the memory and processing capabilities of a single GPU, making distributed execution across multi-GPU clusters essential for both training and inference. As a result, developing methods to efficiently use these distributed systems has become a key challenge in machine learning systems research.

A common approach to managing this complexity is the bulk synchronous parallel~(BSP) programming model, which simplifies development by structuring work into distinct steps: GPUs first perform local computations, then they enter a global communication and synchronization phase. This is typically handled by libraries like AMD ROCm Communication Collectives Library~(RCCL)~\cite{AMD:2025:RCL} or NVIDIA Collective
Communication Library~(NCCL)~\cite{NVIDIA:2025:NCL}, which provide simple function calls for complex communication patterns. This leads to a common execution pattern of ``Compute, Wait, Collective, Wait, Compute.'' While this model is easy to use and has enabled large-scale training and inference, it introduces significant performance inefficiencies.

In this paper, we identify three sources of this inefficiency, which we term the ``Three Taxes.'' These taxes explain the performance gap between the simplified BSP model and the hardware's actual capabilities. The three taxes are:

\begin{itemize}
    \item \textbf{Kernel Launch Overhead Tax}: The cumulative cost of starting separate GPU kernels for each computation and communication stage.
    
    \item \textbf{Bulk Synchronous Tax}: The idle time created when faster GPUs are forced to wait for the slowest GPU at global synchronization points.

    \item \textbf{Inter-Kernel Data Locality Tax}: The performance loss from moving intermediate data to slower off-chip memory between separate computation and communication kernels.
\end{itemize}

We argue that these taxes are not fundamental costs but are artifacts of the rigid BSP programming model. Our proposed solution is to move away from this model by combining, or fusing, computation and communication logic into a single, fine-grained GPU kernel. By treating communication as an integrated part of computation rather than a separate, blocking step, we can create a direct data pipeline, replace global barriers with fine-grained synchronization, and overlap computation with communication. To facilitate this, we use Iris~\cite{Awad:2025:IFM}, a Triton-based library that provides the necessary primitives to implement these fused kernels, allowing for ease of programmability compared to traditional low-level approaches.

The primary contributions of this paper are:

\begin{enumerate}
\item The design and implementation of fine-grained fusing techniques that combine communication and computation primitives within high-level Triton kernels, effectively extending the benefits of single-GPU fused kernels to distributed settings.
    \item An analytical framework, the ``Three Taxes,'' for identifying and characterizing the key inefficiencies of the standard bulk-synchronous model in distributed AI.
   \item  A comprehensive evaluation on both a fundamental building block (All-Gather + General Matrix Multiplication~(GEMM)) and a complex LLM inference kernel (Flash Decode), demonstrating that our approach significantly reduces end-to-end latency by systematically eliminating the three taxes.
   \footnote{Source code available at \href{https://github.com/ROCm/iris}{github.com/ROCm/iris}}
\end{enumerate}

\section{Background and Motivation}

\subsection{GPU Programming with Triton}

Triton is a Python-based language and compiler that enables researchers and developers to write highly efficient, hardware-aware GPU kernels with greater ease than traditional CUDA/HIP programming. It abstracts away much of the low-level complexity while still providing the control needed to achieve high performance. Our work leverages Triton for its programmability and performance.

\subsection{Multi-GPU Communication with Collectives}
To execute tasks in parallel, multiple GPUs must synchronize and exchange data. This is typically managed through collective operations. Popular libraries like RCCL (for AMD GPUs) and NCCL (for NVIDIA GPUs) implement a suite of these collectives. Common examples include:
\begin{itemize}
    \item \textbf{All-Gather}: Each GPU starts with a distinct data shard and ends up with a complete copy of the all the shards gathered from all the participating GPUs (see Figure~\ref{fig:allgather}).

    \begin{figure}[h!]
    \centering
    \includegraphics[width=0.4\textwidth]{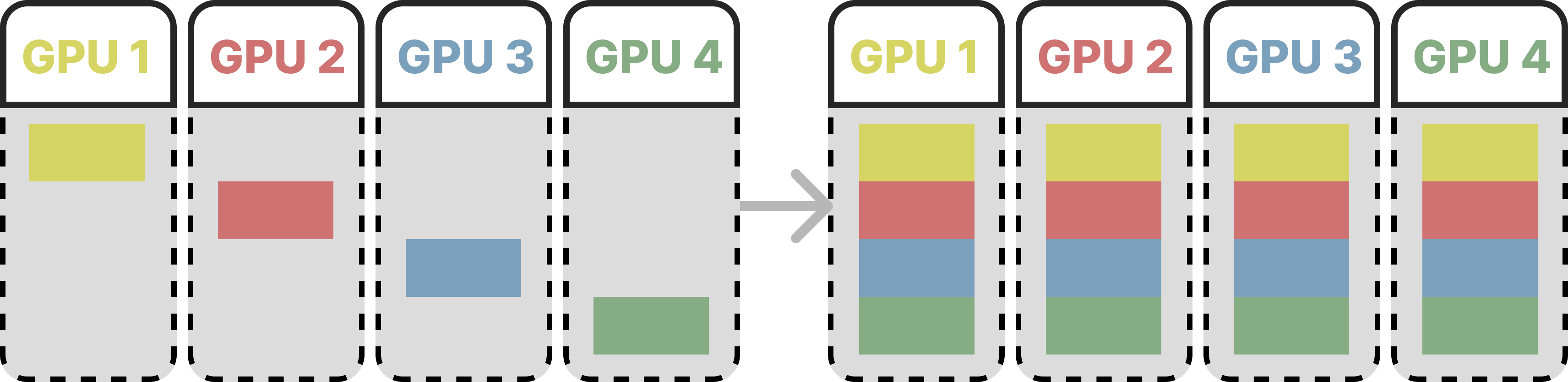}
    \caption{All Gather.}
    \label{fig:allgather}
    \end{figure}
    \item \textbf{All-Reduce}: Data from all GPUs is combined via a specified operation (e.g., sum, max), and the result is shared back such that every participating GPU ends up with a complete copy of the final result.
    
\end{itemize}

\subsection{The Anatomy of Inefficiency: The Three Taxes}
\begin{figure}[h!]
    \centering
    \includegraphics[width=0.4\textwidth]{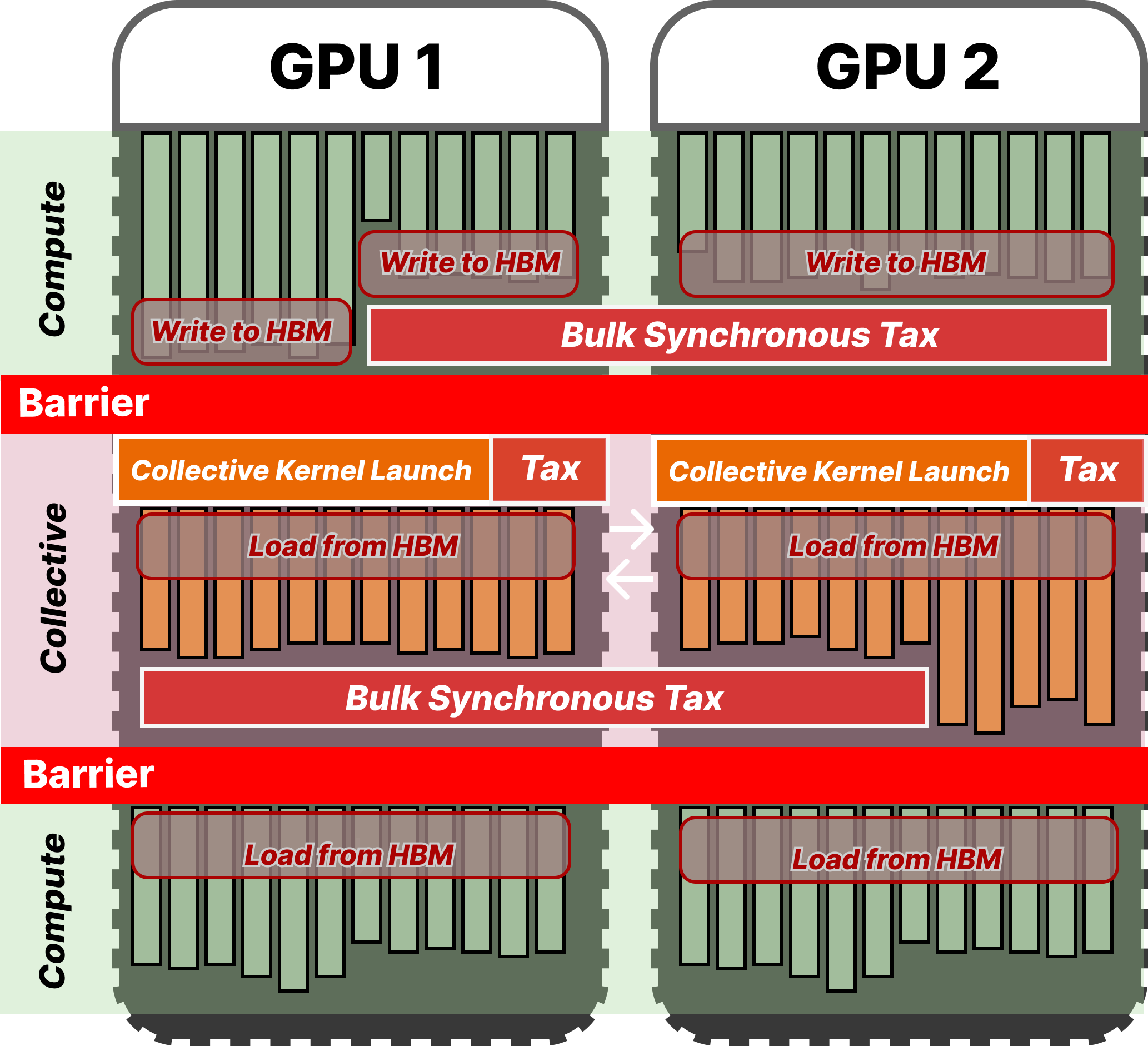}
    \caption{The Three Taxes.}
    \label{fig:taxes}
\end{figure}

The standard pattern when using a library like RCCL is a blocking, five-step process: ``Compute, Wait, Collective, Wait, Compute.'' This can result in a bulk synchronous pattern of execution. While this abstraction simplifies development by providing a clear API call, it introduces systemic inefficiencies. As illustrated in Figure \ref{fig:taxes}, this pattern can be deconstructed into three distinct performance taxes. These taxes represent the performance gap between the simplified bulk synchronous programming model and the hardware's theoretical capabilities.

\subsubsection{Kernel Launch Overhead Tax}
The Kernel Launch Overhead Tax is the fixed latency associated with the dispatch and setup of a GPU kernel by the host. The standard bulk synchronous pattern uses separate, distinct kernel launches for each of the logical steps, for each compute, and each collective call. The cumulative impact of these launches may become a bottleneck in latency-sensitive applications. 

In scenarios where the kernel execution time is short, the kernel launch can constitute a substantial part of the total execution time. Previous work \cite{Spector:2025} has shown that reducing this kernel dispatch time has yielded significant speed-ups. 

\subsubsection{The Bulk Synchronous Tax}

The Bulk Synchronous Tax represents the GPU idle, which results from having the blocking barriers. This tax is paid twice within a single cycle of the ``Compute-Wait-Collective-Wait-Compute'' pattern.

First, after completing the first local computation, all faster GPUs must wait at the first barrier for the slowest GPU before the collective operation can be launched. Second, a synchronization point occurs after the collective operations, where all GPUs must wait again for the data transfer to be fully completed before they can proceed to the next compute kernel.

These global barriers serialize the workflow and create bubbles in the execution pipeline, which lead to a significant under-utilization of hardware resources.

\subsubsection{The Inter-Kernel Tax}
The Inter-Kernel Tax is the cost of breaking the data flow. This is caused by separating the producer and consumer operations into distinct kernels. In the decoupled model, the output of the producer, which resides in the GPU's fast on-chip memory,  is evicted to the slower High-Bandwidth Memory (HBM). Then, the consumer kernel has to bear the full latency and bandwidth cost of fetching the same data from the HBM to begin its operation.
This process breaks the producer-consumer locality. An optimized, fused approach would allow data to be passed directly from the producer to the consumer within the fast cache, avoiding the round-trip to HBM. This tax quantifies the loss of data locality due to the suboptimal data flow between these consecutive operations.

\section{Related Work}
\subsection{ROCm Communication Collectives Library}
Efficient collective communication is essential in distributed and parallel computing, where operations such as all-reduce, broadcast, reduce-scatter, and all-to-all synchronize data across many devices. In large-scale deep learning, these collectives dominate runtime, making optimized implementations critical for scalability. RCCL and NCCL are vendor-optimized frameworks that provide high-performance GPU collectives. RCCL leverages AMD's Infinity Fabric\texttrademark~and network interface cards (NICs) for high-bandwidth, low-latency communication within and across nodes.  RCCL, and similarly NVIDIA's NCCL, exemplify the BSP model in GPU-based collective communication. In this model, collective operations such as All-Reduce, All-Gather, and Reduce-Scatter are initiated via host-side API calls, which launch opaque GPU kernels responsible for executing the communication logic.

We will now unpack the rationale and implications of this design. First, because these kernels are supplied by the vendor, they are opaque to the user and cannot be modified or examined. Second, the BSP model enforces a strict separation between computation and communication phases. As a result, operations must be serialized and independent, which precludes the possibility of constructing producer-consumer pipelines where computation and communication are interleaved with fine-grained, on-device data dependencies.

Such a design choice is motivated by several practical considerations. By abstracting communication into opaque kernels, RCCL ensures portability, stability, and performance consistency across diverse hardware configurations. The BSP model further simplifies synchronization. However, this abstraction comes at the cost of flexibility. The inability to fuse computation and communication within a single kernel limits opportunities for latency hiding and data locality optimization.

Our work addresses this by leveraging Iris, a Triton-based library that provides in-kernel communication primitives, enabling us to write fused kernels where communication and computation logic are combined. This approach enables tile-level producer-consumer pipelines that maintain data locality in on-chip memory and systematically eliminate the Three Taxes.

\subsection{Triton Distributed}
Triton Distributed~\cite{triton-distributed} is a compiler-based extension of OpenAI's Triton compiler~\cite{triton}, developed to address the challenge of overlapping computation and communication in multi-GPU AI deployments. It enables developers to write performant distributed kernels without relying directly on low-level programming models (e.g., HIP/CUDA), instead exposing Python bindings to underlying communication libraries such as rocSHMEM or NVSHMEM\@.

Triton Distributed employes tile-centric primitives to decompose workloads into overlappable segments, enabling fine-grained control over data movement and synchronization. This design facilitates the embedding of communication directly within computational logic, allowing for fused kernel implementations that remove kernel launch overheads and enhance scalability across multi-GPU deployments.

Despite its strengths, Triton Distributed presents several limitations that may impact its usability and extensibility. The framework layers Python bindings over low-level communication libraries such as rocSHMEM and NVSHMEM, but does so by embedding C-style APIs directly into Python and Triton code. This design choice introduces a significant code complexity burden, requiring developers to manually manage thread identifiers and synchronization; this approach ultimately diverges from Triton’s idiomatic programming model and instead relies on Triton anti-patterns~(e.g., manual $tid$). Furthermore, the runtime behaves as a black box, offering limited introspection and control over execution behavior. Triton Distributed also inherits challenges from OpenSHMEM, including the absence of a coherent memory model and unintuitive synchronization primitives, which further complicate reasoning about correctness and performance in distributed kernel development.

\subsection{Iris}
Iris~\cite{Awad:2025:IFM} is a lightweight, Triton-based framework developed by AMD to simplify multi-GPU programming through native support for remote memory access (RMA). Iris introduces clean abstractions and SHMEM-like APIs directly within Triton, enabling developers to write distributed kernels using familiar Pythonic constructs. It provides a full symmetric heap implementation in Python, along with PyTorch-style host APIs for tensor allocation and construction, and Triton-style device APIs for remote load, store, and atomic operations. These features collectively elevate multi-GPU programming to a first-class experience within Triton, preserving its productivity-oriented programming model while enabling fine-grained compute-communication overlap.

Like Triton Distributed, Iris supports GPU-initiated, fine-grained communication phases. This design enables the natural expression of tiled algorithms and fused kernel pipelines, reducing launch overheads and eliminating intermediate buffers. We adopt Iris over Triton Distributed due to its tighter integration with Triton’s idioms and substantially lower code complexity. Unlike Triton Distributed--which embeds low-level C-style interfaces and requires manual thread management--Iris provides high-level, coherent abstractions for expressing multi-GPU programs. Its intuitive APIs, unified memory model, and straightforward synchronization primitives allow developers to implement complex distributed patterns with minimal effort and high performance.

\section{Fused Patterns}
\label{sec:fused-patterns}
To validate and prove the performance of our fused kernels, we have conducted a series of experiments on both fundamental workloads often used in Deep Learning models (All-Gather + GEMM), as well as in complex, real-world LLM inference kernels (Flash Decode). The primary objective was to compare the performance and programmability of the fused implementation compared to a conventional baseline, such as an implementation that uses the RCCL library for communication and execution flow that enforces the bulk synchronous behavior. The experiments were designed to measure end-to-end latency and isolate the performance impact of eliminating the three taxes (Figure \ref{fig:taxes}).

\subsection{All-Gather + GEMM}


\begin{figure}
    \centering
    \includegraphics[width=0.4\textwidth]{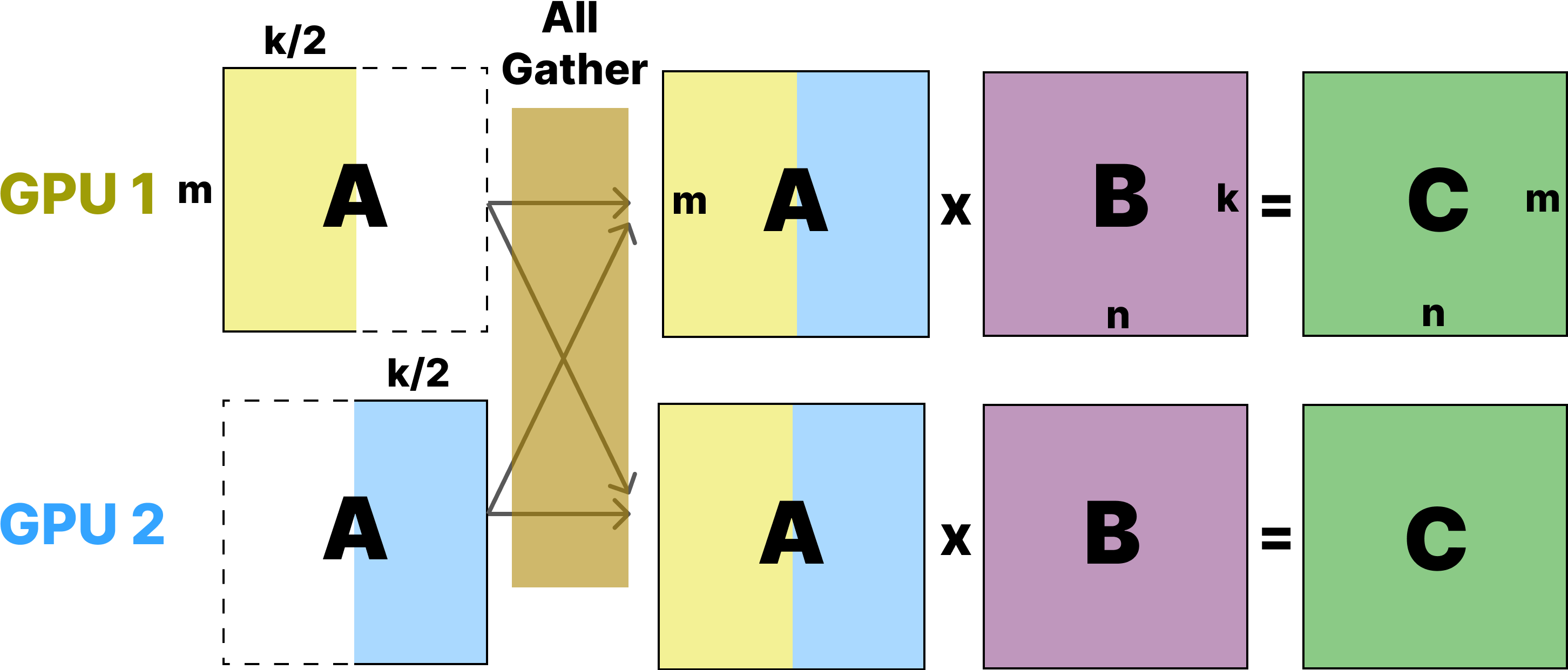}
    \caption{All Gather + GEMM on a Two GPU Setup.}
    \label{fig:ag_gemm_figure}
\end{figure}

\subsubsection{Overview}
We first analyze the All-Gather + GEMM workload (Figure \ref{fig:ag_gemm_figure}), a fundamental building block in numerous distributed ML models. It is often utilized in scenarios like tensor parallelism, where partial results or weights must be collected from all the ranks before a matrix multiply can be performed. 

The operation (Figure \ref{fig:ag_gemm_figure}) consists of a GEMM, $C=A \cdot B$, where the input matrix $A$ is sharded across all GPUs and must first be aggregated via an All-Gather collective before being multiplied with the local matrix $B$.

In our case, the matrix $A$ of size $(M, K)$ was sharded across the $K$ dimension (columns). This results in each $i^\text{th}$ GPU having a matrix $A_i$ of size $(M, K /n)$, where $n$ is the world size. The local matrix $B$ has size $(K, N)$, resulting in the final output matrix $C$ of size $(M,N)$.

We want to note that there are many more possibilities to shard the $A$ and $B$ matrices. We have chosen this particular configuration as it is commonly used in popular LLM frameworks (vLLM, etc.) to support model sharding (Llama, etc.). Other configurations might include: sharding $A$ across the $M$ dimension, sharding $B$ across the $K$ or $N$ dimensions, and any other combination in between.

\subsubsection{Baseline}

The baseline implementation represents the standard, non-fused approach. It involves a blocking \texttt{dist.all\_gather()} call using RCCL to assemble to full matrix A, followed by a \texttt{torch.matmul()} call to perform the computation.

Against this baseline, we evaluated two distinct fine-grained fused strategies, ``Pull'' and ``Push'', which explore different methods of directing the in-kernel data movement and synchronization.

\subsubsection{Pull Model}

\begin{figure}[h!]
    \centering
    \includegraphics[width=0.4\textwidth]{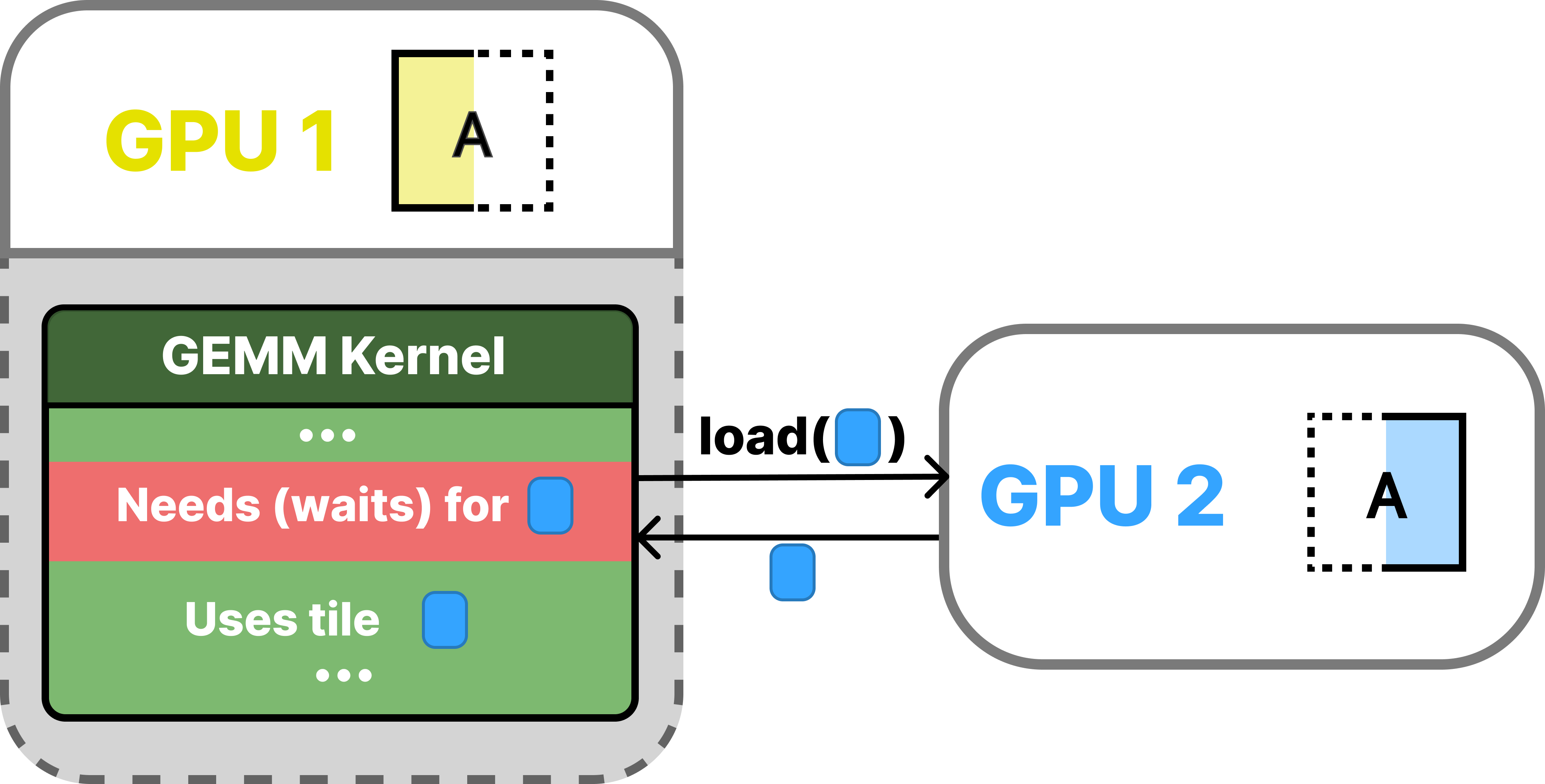}
    \caption{All Gather + GEMM Pull Model.}
    \label{fig:ag_gemm_pull}
\end{figure}

The Pull Model (Figure \ref{fig:ag_gemm_pull}) is a consumer-driven approach where the GEMM kernel takes full responsibility for fetching the data it requires. In this model, we modify the inner loop of a standard Triton GEMM kernel to directly retrieve the remote data on demand.

The implementation (Algorithm \ref{alg:ag_gemm_pull}) replaces the local \texttt{tl.load()} operation for the distributed matrix $A$ with a call to \texttt{iris.load()}, where \texttt{iris.load()} is canonical to Triton's \texttt{tl.load()} with the additional inclusion of remote and local GPU ranks. When the kernel requires a specific tile of $A$ that is located on a remote device, it issues a remote load and implicitly waits for the data to arrive. Synchronization is handled implicitly by the iris command; the compute thread that issued the load will stall until the data is available in its registers, at which point computation resumes. This design integrated data fetching with computation, resulting in a simple implementation, as it does not require explicit synchronization primitives like flags or semaphores. 

\begin{algorithm}[tb]
\caption{Fused All-Gather + GEMM (Pull Model)}
\label{alg:ag_gemm_pull}
\begin{algorithmic}
\STATE {\bfseries Input:} My rank $r$, World size $W$
\STATE {\bfseries Input:} Pointers to remote shards $A_0, A_1, ..., A_{W-1}$
\STATE {\bfseries Input:} Local matrix $B$
\STATE
\STATE Let $C_{tile}$ be the output tile for this thread block
\STATE $acc \gets 0$
\STATE \COMMENT{\textit{Iterate through all GPU shards to perform the All-Gather}}
\FOR{$s \gets 0$ {\bfseries to} $W-1$}
    \FOR{each block $k$ in shard $A_s$}
        \STATE $a_{tile} \gets \text{RemotePull}(A_s(k))$ \COMMENT{\textit{Pull from GPU $s$}}
        \STATE $b_{tile} \gets \text{Load}(B(k))$
        \STATE $acc \gets acc + \text{dot}(a_{tile}, b_{tile})$
        \STATE \COMMENT{\textit{Note: RemotePull is a local copy if $s = r$}}
    \ENDFOR
\ENDFOR
\STATE Store($C_{tile}$, $acc$)
\end{algorithmic}
\end{algorithm}

\subsubsection{Push Model}
\label{sec:push_model}

\begin{figure}[h!]
    \centering
    \includegraphics[width=0.36\textwidth]{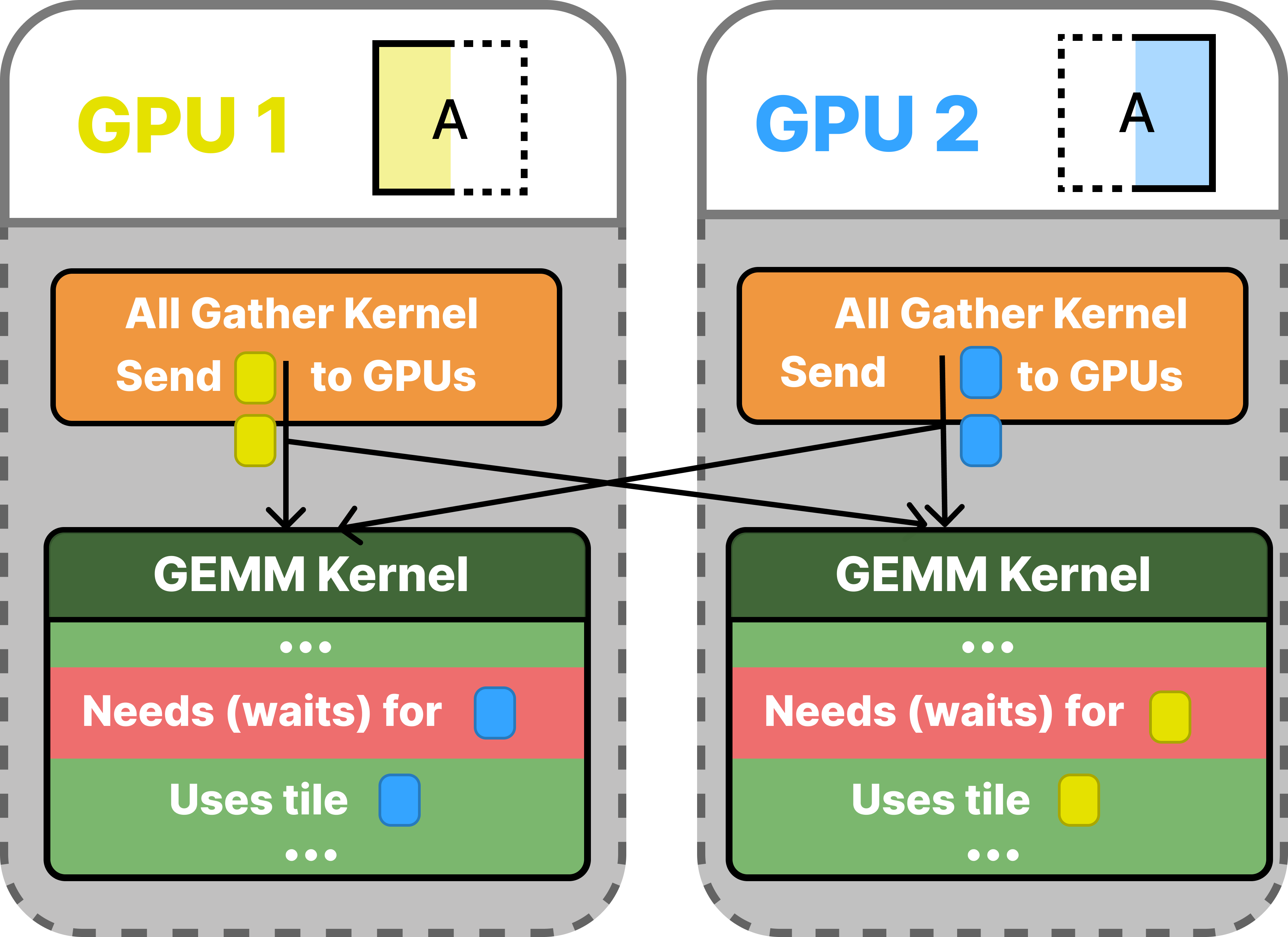}
    \caption{All Gather + GEMM Push Model.}
    \label{fig:ag_gemm_push}
\end{figure}

The Push Model (Figure \ref{fig:ag_gemm_push}) is a producer-driven approach that decouples the communication task into a separate kernel. This model is slightly more complex than the Pull, as it requires explicit synchronization.

The process has two steps:
\begin{enumerate}
    \item A dedicated push kernel (Algorithm \ref{alg:ag_gemm_push_1}) is launched. Its role is to read tiles of the local shard $A_i$ and ``push'' them to the inbox of all remote GPUs using \texttt{iris.store()}, where \texttt{iris.store()} is canonical to Triton's \texttt{tl.store()} with the additional inclusion of remote and local GPU ranks. After a tile is successfully pushed to the destination, the kernel will update a sync flag on the destination GPU that marks the data as ready.
    \item The main GEMM kernel (Algorithm  \ref{alg:ag_gemm_push_2}) executes in parallel. Before computing with a specific tile of $A$, it performs a spin-wait on the corresponding sync flag in its local memory. Once the flag indicates that the remote producer has delivered the data to its inbox, then the compute can resume, and the GEMM kernel will perform a local \texttt{tl.load()} from the inbox to retrieve the tile and proceed with the computation.
\end{enumerate}

This model introduces the overhead of launching a separate push kernel and the complexity of managing explicit synchronization flags.

\begin{algorithm}[tb]
\caption{Fused All-Gather + GEMM (Push Model - Stage 1: Push Kernel)}
\label{alg:ag_gemm_push_1}
\begin{algorithmic}
\STATE {\bfseries Input:} My rank $r$, World size $W$
\STATE {\bfseries Input:} My local shard $A_r$
\STATE {\bfseries Input:} Pointers to remote inboxes $Inbox_0, ..., Inbox_{W-1}$
\STATE {\bfseries Input:} Pointers to remote signal flags $Flags_0, ..., Flags_{W-1}$
\STATE
\STATE \COMMENT{\textit{Broadcast local shard to all other GPUs}}
\FOR{$d \gets 0$ {\bfseries to} $W-1$}
    \FOR{each block $k$ in local shard $A_r$}
        \STATE $a_{tile} \gets \text{Load}(A_r(k))$ \COMMENT{\textit{Load local tile}}
        \STATE $\text{RemotePush}(a_{tile}, \text{Inbox}_d(r, k))$ \COMMENT{\textit{Push to GPU $d$'s inbox}}
        \STATE $\text{RemoteAtomicInc}(Flags_d(r, k))$ \COMMENT{\textit{Signal GPU $d$ block is ready}}
    \ENDFOR
\ENDFOR
\end{algorithmic}
\end{algorithm}

\begin{algorithm}[tb]
\caption{Fused All-Gather + GEMM (Push Model - Stage 2: Wait \& Compute)}
\label{alg:ag_gemm_push_2}
\begin{algorithmic}
\STATE {\bfseries Input:} My rank $r$, World size $W$
\STATE {\bfseries Input:} My inbox $Inbox_r$, My signal flags $Flags_r$
\STATE {\bfseries Input:} Local matrix $B$
\STATE
\STATE Let $C_{tile}$ be the output tile for this thread block
\STATE $acc \gets 0$
\STATE \COMMENT{\textit{Iterate through all GPU shards to perform the All-Gather}}
\FOR{$s \gets 0$ {\bfseries to} $W-1$}
    \FOR{each block $k$ in shard from source $s$}
        \STATE \COMMENT{\textit{Spin-wait until data from GPU $s$ is ready}}
        \WHILE{$Flags_r(s, k) == 0$}
            \STATE pass
        \ENDWHILE
        \STATE $a_{tile} \gets \text{Load}(Inbox_r(s, k))$ \COMMENT{\textit{Load tile from my inbox}}
        \STATE $b_{tile} \gets \text{Load}(B(k))$
        \STATE $acc \gets acc + \text{dot}(a_{tile}, b_{tile})$
    \ENDFOR
\ENDFOR
\STATE Store($C_{tile}$, $acc$)
\end{algorithmic}
\end{algorithm}

\subsection{Flash Decode}

To demonstrate the generalizability and effectiveness of the fine-grained fused patterns on a complex, production-relevant workload, we applied them to Flash Decode \cite{Dao:2024}, a state-of-the-art algorithm for the decoding phase of LLMs. 
Our experimental approach for Flash Decode was to perform a systematic evolution of the implementation. We started with a bulk synchronous baseline, adapted from Triton-Distributed~\cite{triton-distributed}, and progressively integrated our fine-grained techniques in stages. This progression allows us to analyze how each optimization contributes to the final performance gain by systematically tackling the three taxes.

\subsubsection{Overview}

The multi-GPU Flash Decode Algorithm \cite{triton-distributed} can be logically split into three computational stages. First, a kernel computes partial attention scores for the input query against the local shard of the key-value (KV) cache. Second, an ``Online Softmax'' \cite{milakov2018onlinenormalizercalculationsoftmax} kernel computes the normalized partial values based on the local shard. Finally, these partial results must be combined across all GPUs to compute the final, globally correct output. Therefore, an All-Gather collective operation is required to share the locally computed softmax values before the final combine kernel can be executed.

\subsubsection{Flash Decode Baseline}

The Flash Decode baseline implementation is an AMD-adapted version from the Triton-Distributed library, which provides the multi-GPU version of the Flash Decode Algorithm. The baseline follows the Compute-Wait-Collective-Wait-Compute pattern. After the local softmax results are computed, the program makes a call to \texttt{dist.all\_gather()}, using RCCL for the collective. After the collective results are available, the final global combine kernel is launched. Therefore, this baseline is subject to all three taxes: the Kernel Launch, the Bulk Synchronous, and the Inter-Kernel Tax.

\begin{figure}[h!]
    \centering
    \includegraphics[width=0.36\textwidth]{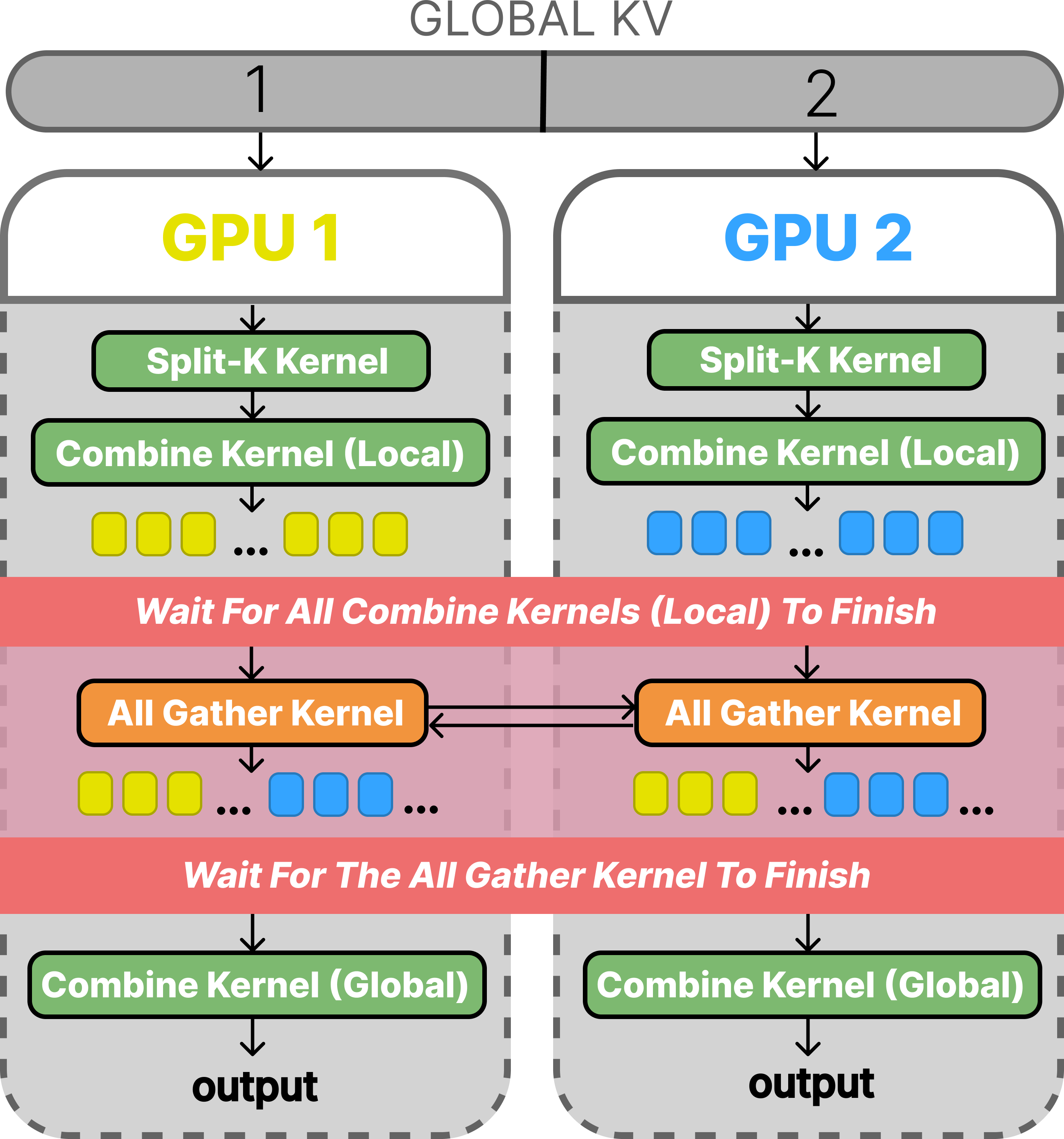}
    \caption{Flash Decode Bulk Synchronous.}
    \label{fig:fd_bsp}
\end{figure}
\subsubsection{Independent All-Gather Kernel}
The first step in our evolving implementation was to replace the opaque RCCL collective with our own functionally equivalent, standalone All-Gather Kernel implemented using Iris. This implementation (Figure \ref{fig:fd_bsp}) is still functionally equivalent to the Triton-Distributed version, which uses their own specific All-Gather Kernel. This Independent All-Gather Kernel still operates in the bulk-synchronous manner; it is launched as a separate kernel that waits for all input data from the previous compute stage, performs the all-to-all data communication, and then waits for the transfer to complete before exiting. While this step allows us to control and fine-tune the communication primitives, it is still subject to the three taxes.

\subsubsection{Fine-Grained Waits}
\begin{figure}[h!]
    \centering
    \includegraphics[width=0.36\textwidth]{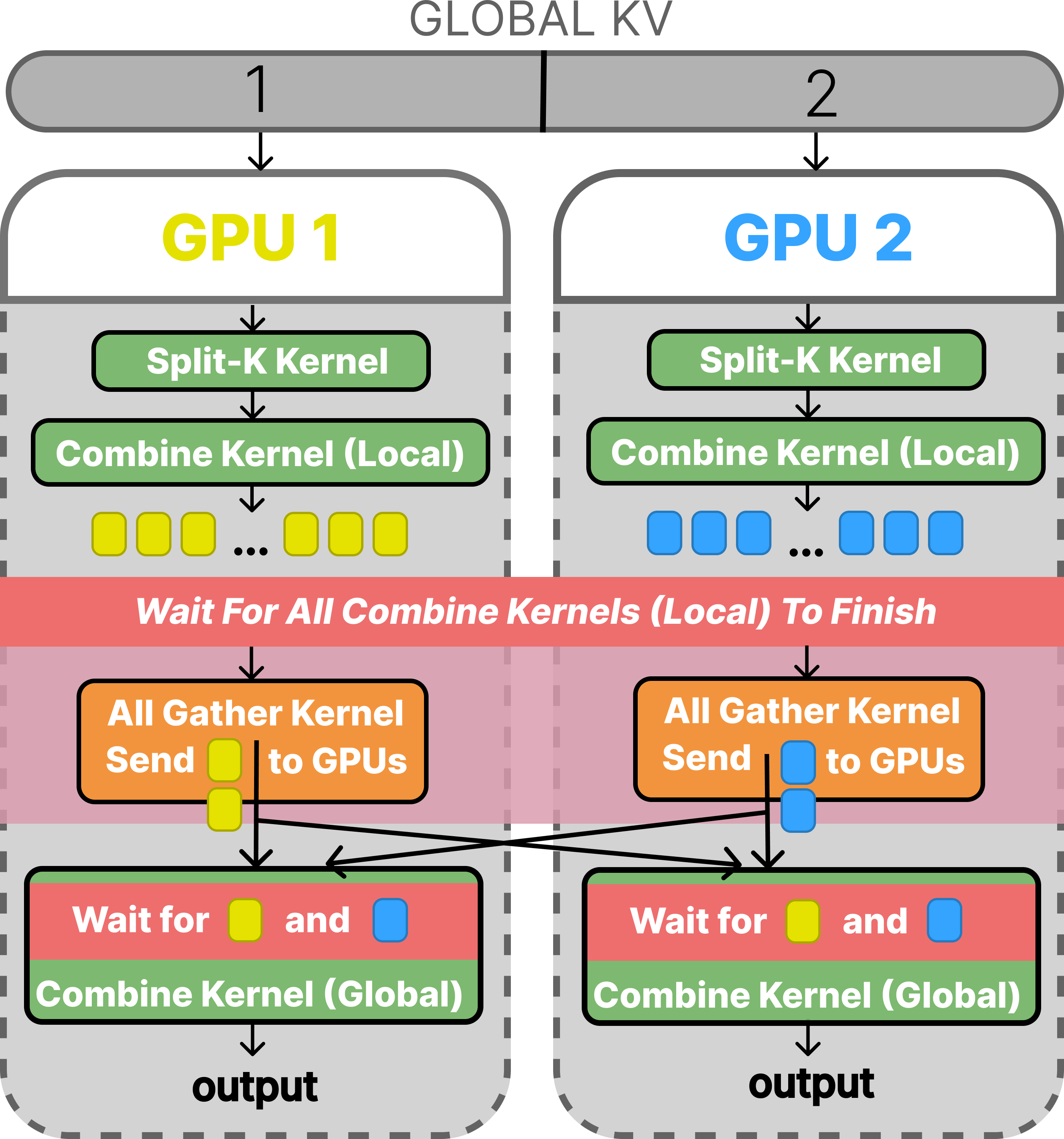}
    \caption{Flash Decode Fine-Grained Waits.}
    \label{fig:fd_wait}
\end{figure}

The second implementation step (Figure \ref{fig:fd_wait}) introduces a change from the bulk synchronous model by focusing on how the data is consumed. The All-Gather Kernel is modified to operate in a non-blocking way from the consumer's perspective. It pushes a tile of data to the remote GPUs and signals their arrival using flags, without waiting for the entire collective to finish.

The consumer kernel (Combine Kernel Global) is then refactored to use fine-grained waits. Instead of a single coarse-grained wait for the whole data, the consumer's inner loop has a spin-wait on the flag for the data it wants to wait for. This allows the consumer kernel to begin processing the first tiles of data as soon as they arrive, overlapping computation with the ongoing communication of later tiles. This approach tackles the Bulk Synchronous Tax from the consumer's perspective, similar to the All-Gather Push variant (Section \ref{sec:push_model}).

\subsubsection{Fused Kernels}

\begin{figure}[h!]
    \centering
    \includegraphics[width=0.36\textwidth]{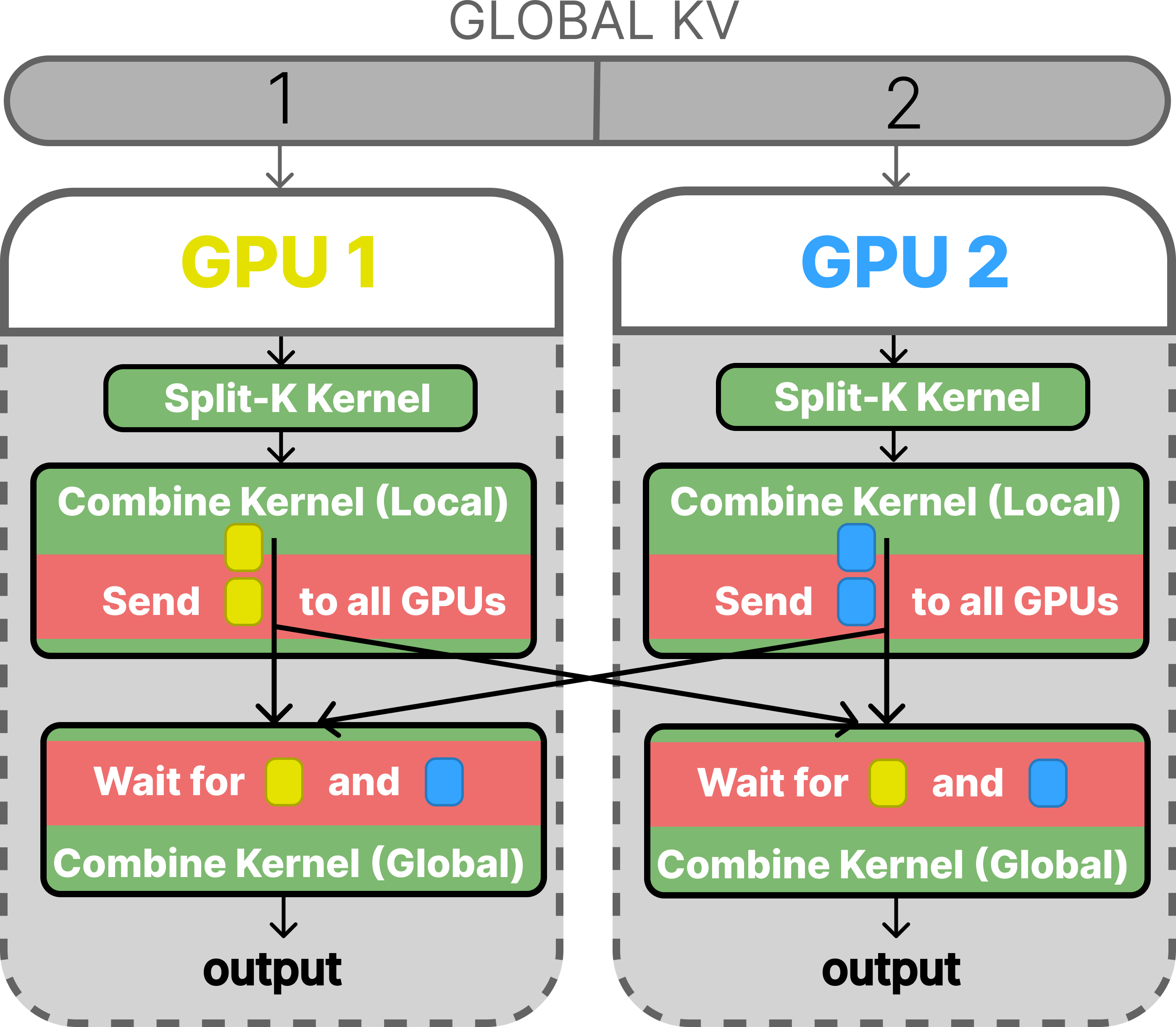}
    \caption{Flash Decode Fused Kernels.}
    \label{fig:fd_fused}
\end{figure}

The final stage (Figure \ref{fig:fd_fused}) is to address how the data is produced, resulting in two Fused Kernels. In this version, we totally eliminate the independent All-Gather kernel. The logic for communicating the data is fused directly into the producer kernel (Combine Kernel (Local)), which computes the local online softmax statistics.

As soon as this kernel computes a tile of the results, it immediately loops through the destination ranks and uses \texttt{iris.store()} to push that tile directly to the inbox of the remote consumer GPUs and signals the availability. This creates the producer-consumer pipeline at a tile level, sending data as soon as it's produced and consuming it as soon as it's ready.

This implementation tackles the Kernel Launch Tax~(i.e., by eliminating the All-Gather Kernel), the Bulk Synchronous Tax, and the Inter-Kernel Tax.

\begin{algorithm}[tb]
\caption{Distributed Flash Decode}
\label{alg:flash_decode}
\begin{algorithmic}
\STATE {\bfseries Input:} My rank $r$, World size $W$
\STATE {\bfseries Input:} Query $Q$, my local KV Cache shard $(K_r, V_r)$
\STATE {\bfseries Input:} Pointers to remote inboxes $Inbox_0, ..., Inbox_{W-1}$
\STATE {\bfseries Input:} Pointers to remote signal flags $Flags_0, ..., Flags_{W-1}$
\STATE

\STATE \COMMENT{\textit{Part 1: Fused Local Attention and Asynchronous Push}}
\STATE $O_r^{partial} \gets \text{Attention}(Q, K_r, V_r)$ \COMMENT{\textit{Compute attention on local KV shard and perform online softmax}}
\FOR{$d \gets 0$ {\bfseries to} $W-1$}
    \STATE $\text{RemotePush}(O_r^{partial}, \text{Inbox}_d(r))$ \COMMENT{\textit{Push partial output to GPU $d$}}
    \STATE $\text{RemoteAtomicInc}(Flags_d(r))$ \COMMENT{\textit{Signal remote GPU that data is ready}}
\ENDFOR
\STATE

\STATE \COMMENT{\textit{Part 2: Concurrent Global Reduction}}
\STATE $O_{final} \gets 0$
\FOR{$s \gets 0$ {\bfseries to} $W-1$}
    \STATE \COMMENT{\textit{Spin-wait for data from GPU $s$}}
    \WHILE{$Flags_r(s) == 0$}
        \STATE pass
    \ENDWHILE
    \STATE $O_s^{partial} \gets \text{Load}(Inbox_r(s))$ \COMMENT{\textit{Load incoming partial output}}
    \STATE $O_{final} \gets \text{Combine}(O_{final}, O_s^{partial})$ \COMMENT{\textit{Reduce via online softmax}}
\ENDFOR
\STATE Store($O_{final}$)
\end{algorithmic}
\end{algorithm}

\section{Evaluation}

\subsection{Experimental setup}
The All-Gather + GEMM experiments were conducted on eight MI325X GPUs, while the Flash Decode experiments were conducted on a single server node equipped with eight AMD Instinct MI300X GPUs, each with 192 GB of HBM3 memory. The GPUs within the node are interconnected with Infinity Fabric\texttrademark, with per GPU aggregate bandwidth of 896GB/s~\cite{amd-mi300x-datasheet}.

The software stack consists of Ubuntu 24.04 in a Docker container with PyTorch 2.6.0 compiled with ROCm 6.4.3 to run the experiments. We used RCCL 2.22.3 for our baseline results. Additionally, the only multi-GPU version of Flash Decode was presented in Triton-Distributed~\cite{triton-distributed}, with an NVIDIA-exclusive implementation. Given our AMD-based system configuration, we initially adapted the code for AMD Instinct MI300X GPUs to enable fused kernel development and facilitate performance evaluation. Our fused kernels were written in Triton~\cite{triton} and leverage the communication primitives from the Iris~\cite{Awad:2025:IFM} library. All kernels were benchmarked using FP16 precision, which is standard for modern LLM workloads.

The primary metric for all experiments is end-to-end latency, measured in milliseconds. To ensure accuracy and consistency, we timed the execution from the host after synchronizing the GPU stream to guarantee all previously submitted work was complete. Each benchmark was run for 500 iterations and averaged out, with an extra 100 runs of warm-up to mitigate any overheads.

\subsection{All-Gather + GEMM}
\begin{figure*}[h!]
    \centering
    \includegraphics[width=0.83\textwidth]{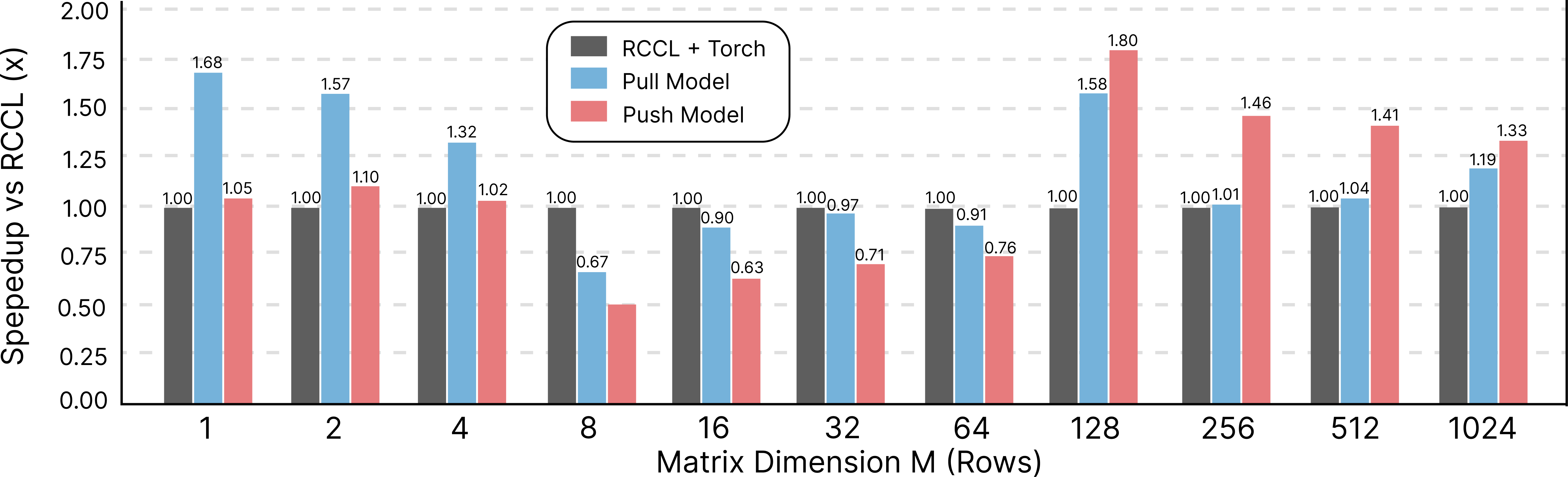}
    \caption{All Gather + GEMM Speedup vs RCCL.}
    \label{fig:ag_gemm_results}
\end{figure*}

The performance of the Pull and Push Models for All-Gather + GEMM workload is presented in Figure \ref{fig:ag_gemm_results}. We use the M, N, K terminology to highlight the shapes of the matrices as discussed in Section~\ref{sec:fused-patterns}. It is using a global N size of 28672, K=8192, running on eight GPUs. We consider the RCCL + Torch implementation as our baseline.

For smaller matrix dimensions $(M \leq 128)$, the Pull model is the more performant approach. By integrating communication directly into the compute kernel, it completely eliminates the Kernel Launch Tax. The Push model, while still tackling the Bulk Synchronous and Inter-Kernel Taxes, must launch a separate kernel for communication, and its associated overhead makes it slower for these short-running workloads.

As the workload size increases $(M \geq 128)$, the Push model becomes the faster of the two. The total execution time is less sensitive to the initial kernel launch cost and more dependent on the raw data movement speed. We have observed that store operations in the Push model are more efficient than the load calls of the Pull model. This superior data movement efficiency outweighs the initial launch overheads, resulting in lower overall latency.

When compared against the RCCL + Torch baseline, our fused kernels are faster at the smallest and largest matrix sizes. However, for configurations where M is between 8 and 64, the baseline is faster than both of our implementations. We believe this is because these specific matrix dimensions are highly optimized in the underlying \texttt{torch.matmul}. For these workloads, the speed of the baseline's GEMM execution appears to be very impactful. Though improved GEMM performance for these sizes would yield improved overall results, we leave further optimization to the GEMM kernel to future work.

\subsection{Flash Decode}

\begin{figure*}[h!]
    \centering
    \includegraphics[width=0.8\textwidth]{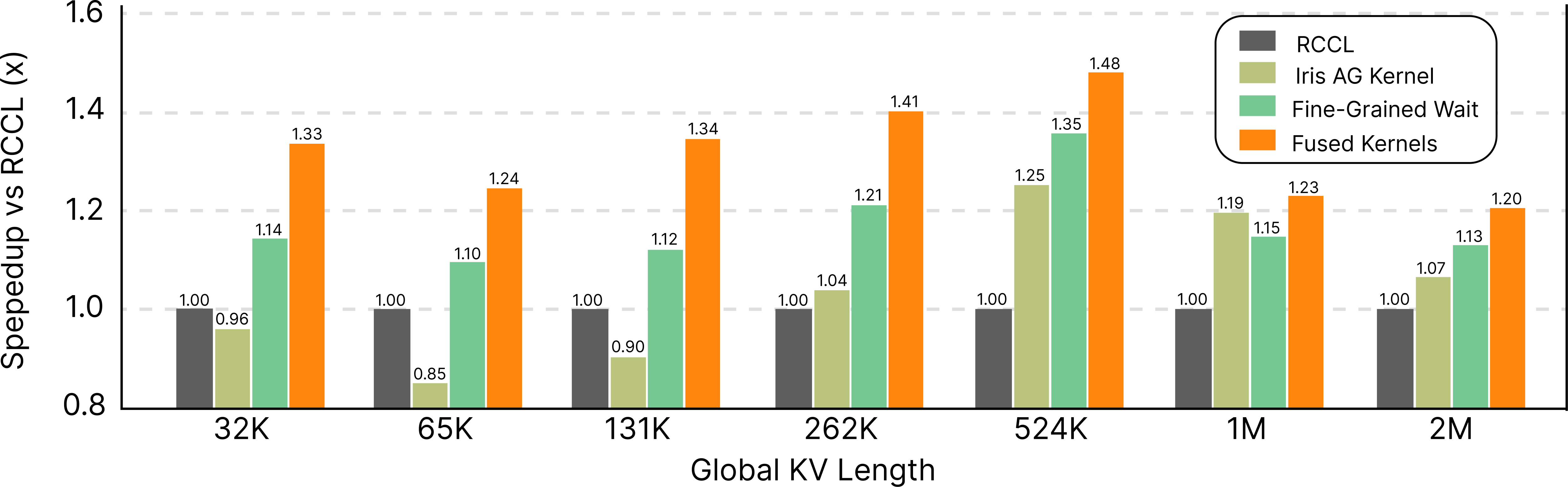}
    \caption{Flash Decode Speedup vs RCCL.}
    \label{fig:fd_latency}
\end{figure*}

The evaluation of the iterative optimizations of Flash Decode is presented in Figure \ref{fig:fd_latency}. The experiments were run with a batch size of one, with 96 query heads, each having a head dimension of 128, and distributed across eight GPUs. Our final fused kernel implementation achieves a 10-20\% speedup compared to the RCCL baseline across a wide range of Global KV Lengths.

The results validate the evolutionary approach:
\begin{itemize}
\item \textbf{Independent AG Kernel vs. RCCL}: The performance of the standalone Iris AG Kernel is very close to the RCCL baseline. This is an expected outcome, as this step replaces the communication library but preserves the bulk synchronous execution model, meaning that it is still subject to the three taxes.

\item \textbf{Fine-Grained Waits}: By introducing Fine-Grained Waits, we can observe a consistent performance improvement over the baseline. This demonstrated the benefit of tackling the consumer side of the Bulk Synchronous Tax, allowing the consumer kernel to process a tile as soon as it arrives.

\item \textbf{Fused Kernels}: This final step eliminates the separate All-Gather Kernel entirely, tackling the Kernel Launch tax. The sender part of the Bulk-Synchronous Tax is also avoided. This cumulative effect of eliminating all three taxes results in the highest performance gains across all sequence lengths.

\end{itemize}

\begin{figure*}[h]
    \centering
    \includegraphics[width=0.68\textwidth]{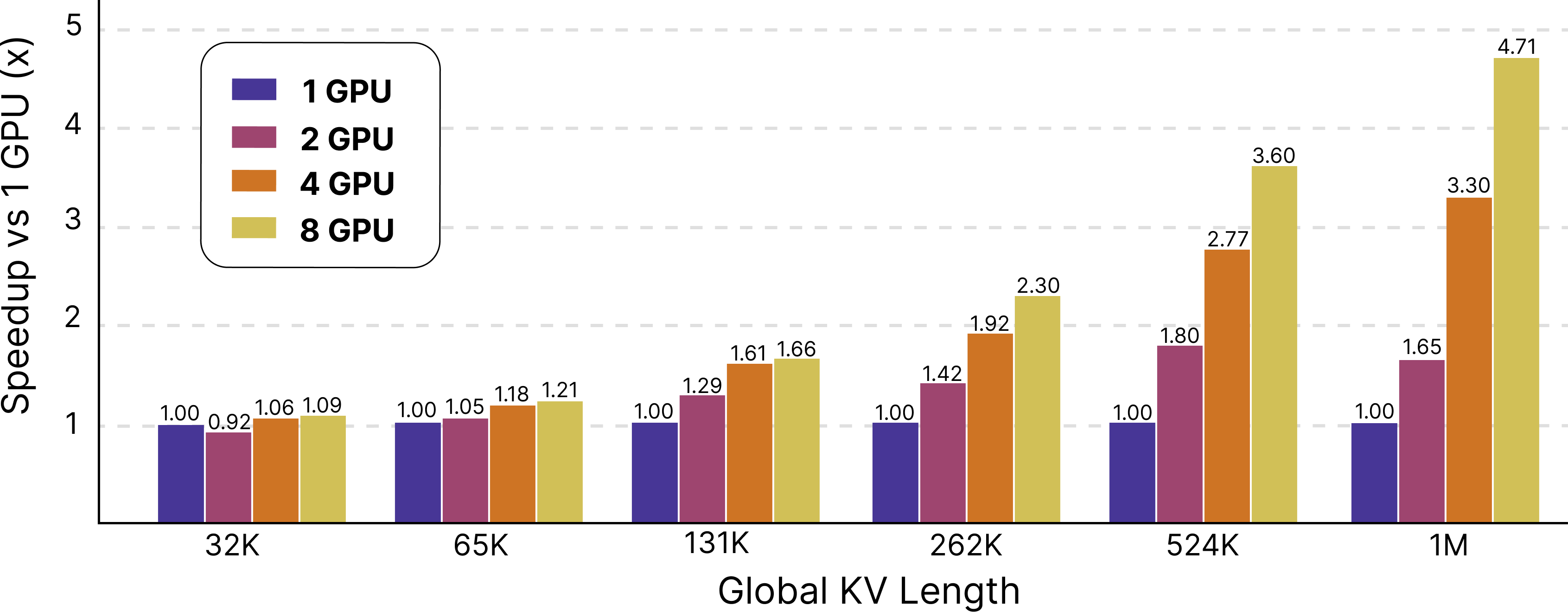}
    \caption{Flash Decode Scaling.}
    \label{fig:gpu_scaling}
\end{figure*}

In addition to comparing against the baseline, we evaluated the scaling properties of the fused implementation for Flash Decode, as shown in Figure \ref{fig:gpu_scaling}. Crucially, as both the workload and GPU count increase, our approach avoids the severe synchronization bottlenecks. By replacing global barriers with a fine-grained dataflow, the implementation productively utilizes additional hardware, which is evident in the substantial reduction in execution time at larger KV lengths as the GPU count is scaled from one to eight. For smaller Global KV Lengths such as 32K, the performance improvement from adding more GPUs is minimal. This is because the workload is not large enough to fully saturate the computational resources, and the benefits of parallelism are offset by the inherent overheads of data distribution. As the problem size increases, the implementation demonstrates strong scaling from 1 to 8 GPUs. While the speedup is not linear due to the inherent costs of inter-GPU communication, these results demonstrate that our fusing methodology provides a strong foundation for efficiently scaling LLM inference for increasingly demanding workloads.

\section{Discussion and Future Work}

Our research demonstrates that by moving away from the rigid Bulk Synchronous Parallel model, we can achieve significant performance gains in distributed LLM kernels. The ``Three Taxes'' framework provides a clear vocabulary for understanding the sources of inefficiency in standard distributed programming, and our fine-grained fused patterns present a practical solution.
\subsection{Programmability}
A key objective of this work was to unlock fine-grained control over communication without forcing developers to abandon high-level programming models for low-level HIP/CUDA. The programmability of our approach is rooted in the design of the Iris library, which integrates multi-GPU communication as a natural extension of the existing Triton programming model.

Iris provides load and store primitives that intentionally mirror the function signatures and semantics of Triton's native memory access functions. This design choice allows a developer to modify an existing compute kernel to handle communication with minimal, intuitive changes. Instead of learning a new, complex API for inter-GPU data movement, a programmer can simply use \texttt{iris.load()} to fetch a remote tile of data where they would have previously used \texttt{tl.load()} for a local one.

This approach significantly lowers the programming complexity typically associated with fine-grained distributed patterns. Oftentimes, patterns can be reused from workload to workload, as we observed from the All Gather + GEMM Push Model to Flash Decode. This makes implementing tile-level producer-consumer pipelines a practical task, enabling the performance gains from eliminating the ``Three Taxes'' without the steep development cost of traditional low-level programming.
\subsection{Generalizability}

We validated our methodology on two workloads at different levels of complexity to demonstrate the broad applicability of our approach. The first, All-Gather + GEMM, is a fundamental building block found in many distributed models, while the second, Flash Decode, is a complex, production-relevant algorithm for LLM inference. The successful application of fine-grained fusing to both a foundational primitive and a state-of-the-art kernel shows that this is a widely generalizable technique.

The core of our method relies on the ``Three Taxes'' analytical framework, which is not specific to any single kernel. This framework can be used to identify the same inherent inefficiencies (kernel launch overhead, bulk synchronous waits, and inter-kernel data movement) in any algorithm that follows the standard ``Compute-Wait-Collective-Wait-Compute'' pattern.

Therefore, the fused patterns can be generalized to other collective operations common in deep learning. For instance, training workloads could benefit from fusing Reduce-Scatter or All-Reduce operations directly. The primary requirement for this approach to be effective is that the workload can be decomposed into smaller, tile-level operations, which creates the opportunity to overlap computation with fine-grained communication.
\subsection{Future Work}

Our methodology for fusing computation and communication is highly generalizable and can be extended beyond the specific kernels presented in this work. The next logical step is to apply these fine-grained fused patterns to a broader range of LLM workloads, including the performance-critical stages of inference and various training kernels. Developing fused versions of the whole LLM life cycle would be a significant contribution toward a comprehensive library of high-performance distributed primitives.

This fusing of communication and computation into a single scope unlocks a powerful new optimization capability. By bringing communication parameters, such as the granularity of data transfer, into the same kernel as computation parameters like tile size, we can leverage a unified autotuning approach. Triton's existing autotuner could be extended to explore this combined search space, simultaneously optimizing for both computation and communication to find the true optimal configuration. This is a significant departure from the traditional method of tuning these aspects in isolation and has the potential to maximize hardware utilization and latency-hiding in a way that was not previously possible.

\section{Conclusion}
This paper addresses the performance inefficiencies of the Bulk Synchronous Parallel model, a common bottleneck in distributed AI workloads, particularly Large Language Models. We introduced the ``Three Taxes'' (Kernel Launch Overhead, Bulk Synchronous, and Inter-Kernel Data Locality) as an analytical framework to characterize the bottlenecks inherent in the standard ``Compute-Wait-Collective-Wait-Compute'' pattern. Our solution systematically eliminates these taxes by fusing communication logic directly into computation kernels, a task made programmable through the Iris library for Triton.

By creating fine-grained, tile-level producer-consumer pipelines, we replace coarse-grained global barriers with efficient dataflow synchronization. We validated this approach on two critical and distinct workloads: the foundational All-Gather + GEMM operation and the complex Flash Decode algorithm used in LLMs. Across a wide range of problem sizes, our fused kernels demonstrated a 10-20\% speedup in end-to-end latency over the RCCL baselines. This work establishes a more programmable and efficient paradigm for large-scale distributed AI, demonstrating that significant performance gains can be unlocked by moving beyond the rigid BSP model.

\nocite{langley00}

\bibliography{example_paper}

\begin{thebibliography}{9}
\providecommand{\natexlab}[1]{#1}
\providecommand{\url}[1]{\texttt{#1}}
\expandafter\ifx\csname urlstyle\endcsname\relax
  \providecommand{\doi}[1]{doi: #1}\else
  \providecommand{\doi}{doi: \begingroup \urlstyle{rm}\Url}\fi

\bibitem[{Advanced Micro Devices, Inc.}(2024)]{amd-mi300x-datasheet}
{Advanced Micro Devices, Inc.}
\newblock {AMD Instinct MI300X Platform Data Sheet}.
\newblock \url{https://www.amd.com/content/dam/amd/en/documents/instinct-tech-docs/data-sheets/amd-instinct-mi300x-platform-data-sheet.pdf}, 2024.
\newblock Accessed: 2025-10-17.

\bibitem[{AMD}(2025)]{AMD:2025:RCL}
{AMD}.
\newblock {RCCL}: {ROCm} communication collectives library, July 2025.
\newblock URL \url{https://github.com/ROCm/rccl}.
\newblock [Online; accessed 27-October-2025].

\bibitem[Awad et~al.(2025)Awad, Osama, and Potter]{Awad:2025:IFM}
Awad, M., Osama, M., and Potter, B.
\newblock {Iris}: First-class multi-{GPU} programming experience in {Triton}, October 2025.
\newblock URL \url{https://github.com/ROCm/iris}.

\bibitem[Dao et~al.(2024)Dao, Haziza, Massa, and Sizov]{Dao:2024}
Dao, T., Haziza, D., Massa, F., and Sizov, G.
\newblock Flash-decoding for long-context inference, 2024.
\newblock URL \url{https://crfm.stanford.edu/2023/10/12/flashdecoding.html}.

\bibitem[Milakov \& Gimelshein(2018)Milakov and Gimelshein]{milakov2018onlinenormalizercalculationsoftmax}
Milakov, M. and Gimelshein, N.
\newblock Online normalizer calculation for softmax, 2018.
\newblock URL \url{https://arxiv.org/abs/1805.02867}.

\bibitem[{NVIDIA}(2025)]{NVIDIA:2025:NCL}
{NVIDIA}.
\newblock {NCCL}: Optimized primitives for collective multi-{GPU} communication, July 2025.
\newblock URL \url{https://github.com/NVIDIA/nccl}.
\newblock [Online; accessed 27-October-2025].

\bibitem[Spector et~al.(2025)Spector, Juravsky, Sul, Dugan, Lim, Fu, Arora, and Ré]{Spector:2025}
Spector, B., Juravsky, J., Sul, S., Dugan, O., Lim, D., Fu, D., Arora, S., and Ré, C.
\newblock Look ma, no bubbles! designing a low-latency megakernel for llama-1b, May 2025.
\newblock URL \url{https://hazyresearch.stanford.edu/blog/2025-05-27-no-bubbles}.

\bibitem[Tillet et~al.(2019)Tillet, Kung, and Cox]{triton}
Tillet, P., Kung, H.~T., and Cox, D.
\newblock Triton: an intermediate language and compiler for tiled neural network computations.
\newblock In \emph{Proceedings of the 3rd ACM SIGPLAN International Workshop on Machine Learning and Programming Languages}, MAPL 2019, pp.\  10–19, New York, NY, USA, 2019. Association for Computing Machinery.
\newblock ISBN 9781450367196.
\newblock \doi{10.1145/3315508.3329973}.
\newblock URL \url{https://doi.org/10.1145/3315508.3329973}.

\bibitem[Zheng et~al.(2025)Zheng, Bao, Hou, Zheng, Fang, Huang, Li, Duanmu, Chen, Xu, Guo, Zheng, Jiang, Di, Wang, Ye, Lin, Chang, Lu, Liang, Zhai, and Liu]{triton-distributed}
Zheng, S., Bao, W., Hou, Q., Zheng, X., Fang, J., Huang, C., Li, T., Duanmu, H., Chen, R., Xu, R., Guo, Y., Zheng, N., Jiang, Z., Di, X., Wang, D., Ye, J., Lin, H., Chang, L.-W., Lu, L., Liang, Y., Zhai, J., and Liu, X.
\newblock Triton-distributed: Programming overlapping kernels on distributed ai systems with the triton compiler, 2025.
\newblock URL \url{https://arxiv.org/abs/2504.19442}.

\end{thebibliography}
\bibliographystyle{mlsys2025}

%


\end{document}